\documentclass[english,prd, 11pt, nofootinbib,  superscriptaddress, preprintnumbers,floatfix]{revtex4}
\usepackage[T1]{fontenc}
\usepackage[latin9]{inputenc}
\usepackage{babel}
\usepackage{mathrsfs}
\usepackage{amsmath}
\usepackage{amssymb}
\usepackage{graphicx}
\usepackage{hyperref}

\providecommand{\tabularnewline}{\\}

\usepackage{slashed}
\usepackage{bbold}
\usepackage{multirow}

\usepackage{feynmp-auto}
\usepackage{xfrac}

\usepackage{amsfonts}
\usepackage{makecell}
\usepackage[dvipsnames]{xcolor}
\usepackage{ulem}
\usepackage{longtable}
\usepackage{array}
\usepackage{subfigure}

\newcommand{\eq}{\begin{eqnarray}}
\newcommand{\en}{\end{eqnarray}}

\newcommand{\corwu}[2]{#2}

\begin{document}
\title{$DDK$ system in  finite volume}
\author{Jin-Yi Pang }
\address{College of Science, University of Shanghai for Science and Technology,
Shanghai 200093, China}
\author{Jia-Jun Wu\footnote{Corresponding author. wujiajun@ucas.ac.cn}}
\address{School of Physical Sciences, University of Chinese Academy of Sciences,
Beijing 100049, China}
\author{Li-Sheng Geng\footnote{Corresponding author. lisheng.geng@buaa.edu.cn}}
\affiliation{School of Physics, Beihang University, Beijing 100191, China}
\affiliation{
Beijing Key Laboratory of Advanced Nuclear Materials and Physics,
Beihang University, Beijing 100191, China}
\affiliation{Beijing Advanced Innovation Center for Big Data-Based Precision Medicine, School of Medicine and Engineering, Beihang University, Beijing, 100191}
\affiliation{School of Physics and Microelectronics, Zhengzhou University, Zhengzhou, Henan 450001, China}

\begin{abstract}
The $DDK$ 3-body system is supposed to be bound due to the strongly attractive interaction between the $D$ meson and the $K$ meson in the isospin zero channel.
The minimum quark content of this 3-body bound state is $cc\bar{q}\bar{s}$ with $q=u,d$.
It will be an explicitly exotic tetraquark state once discovered.
In order to confirm the phenomenological study of the $DDK$ system, we can refer to lattice QCD as a powerful theoretical tool parallel to the experiment measurement.
In this paper, a 3-body quantization condition scheme is derived via the non-relativistic effective theory and the particle-dimer picture in finite volume.
Lattice spectrum of this 3-body system is calculated within the existing model inputs.
The spectrum shows various interesting properties of the $DDK$ system, and it may reveal the nature of the $D^*(2317)$.
This predicated spectrum is expected to be tested in future lattice simulations.
\end{abstract}
\maketitle

\section{Introduction}

The discovery of the $D_{s0}^{*}(2317)$ \cite{Aubert:2003fg,Besson:2003cp,Krokovny:2003zq}
and subsequent phenomenological studies \cite{Bardeen:2003kt,Nowak:2003ra,vanBeveren:2003kd,Dai:2003yg,Narison:2003td,Szczepaniak:2003vy,Browder:2003fk,Barnes:2003dj,Cheng:2003kg,Chen:2004dy,Dmitrasinovic:2005gc,Zhang:2018mnm,Terasaki:2003qa,Maiani:2004vq}
imply that there is a strongly attractive interaction between $D$
and $K$.
The prevailing idea that the $D_{s0}^{*}(2317)$ can be understood as a molecular state has been confirmed by many Weinberg-Tomozawa potential calculations \cite{Kolomeitsev:2003ac,Hofmann:2003je,Guo:2006fu,Gamermann:2006nm,Guo:2008gp,Guo:2009ct,Cleven:2010aw,Yao:2015qia,Guo:2015dha,Albaladejo:2016lbb,Du:2017ttu,Altenbuchinger:2013vwa,Altenbuchinger:2013gaa,Albaladejo:2018mhb,Geng:2010vw,Wang:2012bu,Liu:2009uz}
and lattice QCD simulations \cite{MartinezTorres:2011pr,Guo:2018kno,Liu:2012zya,Guo:2018ocg,Guo:2018tjx,Mohler:2013rwa,Lang:2014yfa,Torres:2014vna,Bali:2017pdv}.
It is also found that there is a long-range attractive potential between $D$ and $D_{s0}^{*}(2317)$. This potential can be explained by the exchange of a nearly on-shell $K$ meson \cite{SanchezSanchez:2017xtl}.
Subsequently, it is very natural to ask what happens in the $DDK$ three-body system.
Till now, many works \cite{SanchezSanchez:2017xtl,MartinezTorres:2018zbl,Wu:2019vsy,Huang:2019qmw} have been devoted to studying this problem. The reference \cite{Wu:2019vsy} found that a $DDK$ bound state exists with a binding energy of about
$70$\,\text{MeV} due to the attractive $DK$ interaction (the repulsive $DD$ interaction  is also considered).
The quantum numbers are $J^{P}=0^{-}$, $I=1/2$, $S=1$ and $C=2$.
It turns out that the minimum quark content of this state is $cc\bar{q}\bar{s}$ with $q=u,d$.
Therefore, once this state is discovered, it should be an explicitly exotic tetraquark state.

Lattice QCD provides an alternative tool to understand considerable hadronic processes based on the QCD first principle.
Recently, lattice simulations have started to probe  three-body systems \cite{Horz:2019rrn,Mai:2019fba,Culver:2019vvu,Blanton:2019vdk}.
A lot of progress has been made in analyzing of these three-body lattice data \cite{Polejaeva:2012ut,Meissner:2014dea,Guo:2016fgl,Guo:2017ism,Briceno:2012rv,Hansen:2014eka,Hansen:2015zga,Hansen:2015zta,Hansen:2016fzj,Hansen:2016ync,Briceno:2017tce,Kreuzer:2010ti,Kreuzer:2008bi,Kreuzer:2009jp,Kreuzer:2012sr,Sharpe:2017jej,pang1,pang2,Meng:2017jgx,pang3,Mai:2017bge,Briceno:2018mlh,Romero-Lopez:2018rcb,Mai:2018djl,Guo:2018xbv,Briceno:2018aml,Guo:2018ibd,Blanton:2019igq,pang4,Hansen:2020zhy,Romero-Lopez:2019qrt} (for the recent review, see Ref. \cite{Hansen:2019nir}).
The finite volume spectrum produced by lattice simulations
can help determine the low energy constants (LECs) in effective field theory and further test the phenomenological works.
The references \cite{pang1,pang2,pang3,pang4} supply a 3-body quantization condition scheme via the non-relativistic effective theory and the particle-dimer picture in  finite volume.
This method can be applied in many hadronic processes transparently and connect lattice QCD simulations with established phenomenological models.

In the current paper, we focus on the $DDK$ system and apply the phenomenological model \cite{Wu:2019vsy} in the finite volume formalism of \cite{pang2} to calculate the corresponding spectrum.
This work starts from $DK$ and $DD$ 2-body scattering, then builds a cut-off independent 3-body formalism of $DDK$ system,
aiming to obtaining the energy levels in  finite volume in the end.
This spectrum is automatically divided into three parts, above the $DDK$ threshold, below the $DD_s^*(2317)$ threshold, and between these two thresholds.
Such spectrum provides various interesting physical information of the $DDK$ system.

The paper is organized as follows. In Sec. \ref{sec:Formalism}, we give the basic formalism to describe the 3-body system of $DDK$ and derive the quantization condition.
In Sec. \ref{sec:Result}, the matching of the 2-body scattering amplitude including $DK$ and $DD$ channels is given based on the setup of Refs.~\cite{Wu:2019vsy,Liu:2019stu}.
We build a 3-body force to generate the $DDK$ bound state predicted by \cite{Wu:2019vsy} and use quantization condition to produce its lattice spectrum.
The calculation up to $O(p^{0})$, $O(p^{2})$, and $O(p^{4})$ are compared to guarantee the stability of the non-relativistic framework.
Finally, the \corwu{discussion}{summary} and outlook are given in Sec. \ref{sec:Discussion}.

\section{\label{sec:Formalism}Formalism}

\subsection{Particle-dimer formalism}

We derive the 3-body quantization condition for the $DDK$ system in the particle-dimer formalism used in \cite{pang1,pang2,pang3}.
In the particle-dimer picture, the $DDK$ system is described by a $DK$-dimer with the other $D$ meson as a spectator.
In principle, the dimer $DK$ can give all the 2-body dynamics in the $DK$ sub-system, not necessarily constrained as a 2-body bound state or resonance, e.g., the $D_{s0}^{*}(2317)$.
To complete a full physical picture, we also need to consider the $DD$-dimer with the $K$ meson as the spectator.
Therefore, the effective Lagrangian reads,
\begin{align}
\mathcal{L}=\, & \mathcal{L}_{1}+\mathcal{L}_{2}+\mathcal{L}_{3},\label{eq:dimer-L}
\end{align}
where
\begin{align}
\mathcal{L}_{1}=\, & D^{\dagger}(i\partial_{0}+\dfrac{\nabla^{2}}{2m_{D}})D+K^{\dagger}(i\partial_{0}+\dfrac{\nabla^{2}}{2m_{K}})K+T_{DK}^{\dagger}\sigma_{DK}T_{DK}+T_{DD}^{\dagger}\sigma_{DD}T_{DD};\\
\mathcal{L}_{2}=\, & T_{DK}^{\dagger}\left[D\mathscr{F}_{DK}K\right]+T_{DD}^{\dagger}\left[D\mathscr{F}_{DD}D\right]+\text{h.c.},\quad\text{where the operator reads, }\mathscr{F}=f_{0}+f_{2}\overleftrightarrow{\nabla}^{2};\\
\mathcal{L}_{3}=\, & h_{0}\left[T_{DK}^{\dagger}D^{\dagger}\right]\left[T_{DK}D\right].
\end{align}
In the Lagrangian, we have kinematic part $\mathcal{L}_1$ including both single particle fields and dimer fields, saying $D$, $K$ and $T_{DD}$, $T_{DK}$.  The dynamics of 2-body sector is given in $\mathcal{L}_2$ where the operators encode the interactions between the dimer fields and their constituents, e.g., $DK$-dimer and $D$ meson, $K$ meson. In the formalism, the 2-body LECs are denoted by $f_{0}$ and $f_{2}$. The remaining interaction is 3-body contact term induced by the dimer and a spectating single particle, e.g., $DK$-dimer and another $D$ meson just as we have shown in $\mathcal{L}_3$. Here, for all the possible 3-body contact interaction, we choose only one operator for the channel $(DK)+D\to(DK)+D$ and a corresponding LEC, i.e., $h_0$. 
In principle, the contact term can also appear in the interaction between the $DD$-dimer and the spectating $K$ meson, as well as the crossed channel $(DK)+D\to(DD)+K$ or the inverse case.
It seems  to be more complete that we parameterize 3-body force in the form of,
\begin{align}
h_{0}\left(T_{DK}^{\dagger}D^{\dagger}\right)\left(T_{DK}D\right)
+h_{0}^{'}\left(T_{DD}^{\dagger}K^{\dagger}\right)\left(T_{DD}K\right)
+h_{0}^{''}\left(\left(T_{DD}^{\dagger}K^{\dagger}\right)\left(T_{DK}D\right)+\text{h.c.}\right)
& .
\end{align}
As a matter of fact, considering the equivalent effective field theory without dimers\footnote{The
equivalence can be seen by integrating the dimer fields out in Lagrangian
(\ref{eq:dimer-L}).}, we have only one operator to build up the 3-body
interaction for the $DDK$ system, that is $h\left(D^{\dagger}D^{\dagger}K^{\dagger}\right)\left(DDK\right)$.
Consequently, the relationship between the general LEC, $h$ and the LECs, $h_{0},h_{0}^{'}, h_{0}^{''}$ in particle-dimer formalism is
\begin{align}
h=\, & \frac{f_{0,DK}^{2}}{\sigma_{DK}^{2}}h_{0}+\frac{f_{0,DD}^{2}}{\sigma_{DD}^{2}}h_{0}^{'}
+\frac{2f_{0,DD}f_{0,DK}}{\sigma_{DK}\sigma_{DD}}h_{0}^{''}.
\end{align}
Since dimer fields are, nevertheless, auxiliary fields, physical observables
can only determine $h$, but have nothing to do with the distribution
of $h_{0}$, $h_{0}^{'}$ and $h_{0}^{''}$,
we are allowed to choose $h_{0}^{'}=h_{0}^{''}=0$ and use the contact term in $(DK)+D\to(DK)+D$ to parameterize the 3-body force.

There are many other ways to parameterize the particle-dimer formalism, for example, using the $N/D$ method with the analytic properties derived from 3-body unitarity \cite{Mai:2017bge}.
Nevertheless, all these parameterizations  give the same 3-body quantization condition in  finite volume.

\subsection{Particle-dimer scattering equation}

The scattering amplitude of $DDK\to DDK$ encodes all the information of the 3-body system. In infinite volume, we can introduce the particle-dimer scattering equation to resolve the analytic properties of the amplitude.
The equation reads,
\begin{align}
\mathcal{M}(\mathbf{p},\mathbf{q};E)= & Z(\mathbf{p},\mathbf{q};E)+4\pi\int^{\Lambda}\frac{d^{3}k}{(2\pi)^{3}}Z(\mathbf{p},\mathbf{k};E)\tau(\mathbf{k};E)\mathcal{M}(\mathbf{k},\mathbf{q};E).\label{eq:scattering-eq}
\end{align}
$\mathcal{M}$ denotes the particle-dimer scattering amplitude which resolves the 3-body information equivalently. Since there are two kinds of dimers, $\mathcal{M}$ takes the form of $2\times2$ matrix
as
\begin{align}
\mathcal{M}= & \begin{pmatrix}\mathcal{M}_{1} & \mathcal{M}_{12}\\
\mathcal{M}_{21} & \mathcal{M}_{2}
\end{pmatrix}.
\end{align}
The subscript $1$ represents the channel of scattering between dimer $DK$ and spectating $D$ meson while $2$ means the channel of $DD$ and $K$.
The off-diagonal term is thus the amplitude of the corresponding cross channel.
We introduce a momentum cutoff $\Lambda$ for the spectating particle to treat UV divergence.

\begin{figure}
\begin{centering}
\includegraphics[scale=0.3]{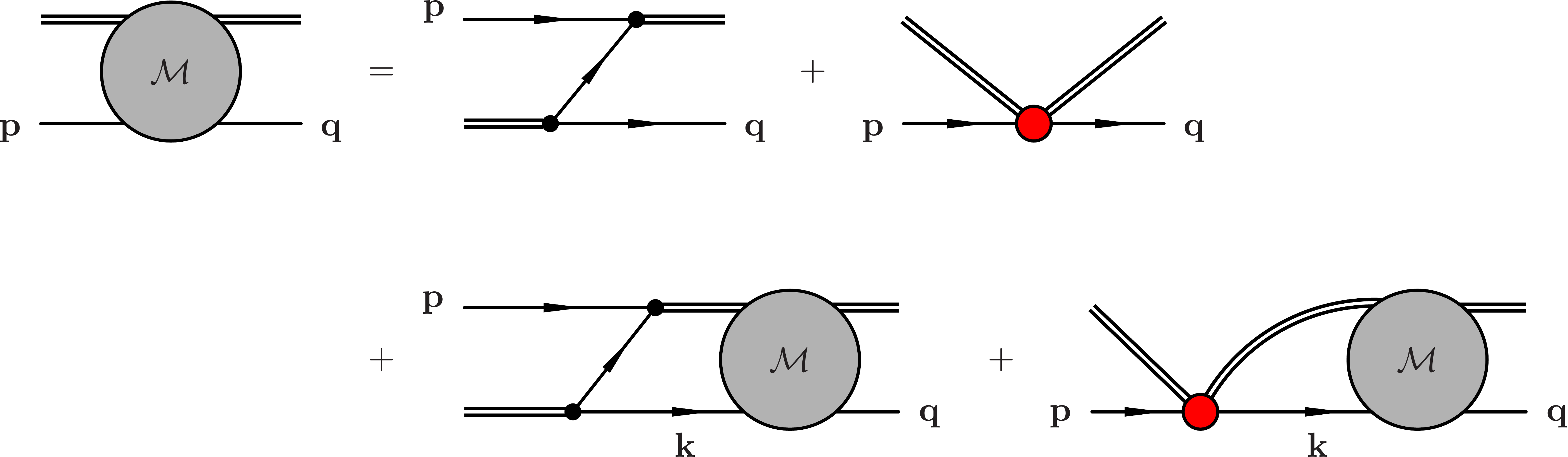}
\par\end{centering}
\caption{Schematic 3-body scattering equation.}
\end{figure}
In the 2-body sector, $\tau$ stands for the dressed dimer propagators in two channels,
\begin{align}
\tau= & \begin{pmatrix}\tau_{1}\\
 & \tau_{2}
\end{pmatrix}.
\end{align}
We can write down the explicit forms for these propagators within non-relativistic kinematics and $S$-wave projection,
\begin{align}
\tau_{1}(\mathbf{k};E)=\frac{1}{k_{*,1}\cot\delta_{DK}-ik_{*,1}} & ,\quad k_{*,1}^{2}=2\mu E-\beta^{2}k^{2};\\
\tau_{2}(\mathbf{k};E)=\frac{2Z_{DD}}{k_{*,2}\cot\delta_{DD}-ik_{*,2}} & ,\quad k_{*,2}^{2}=2\mu E/Z_{DD}-\alpha^{2}k^{2}.
\end{align}
where $\mu$ is the \corwu{2-body reduced mass}{$DK$ 2-body reduced mass, i.e., $\mu=\dfrac{m_Km_D}{m_K+m_D}$,}  $\gamma=\dfrac{\mu}{m_{K}}$, $\beta^{2}=1-\gamma^{2}$, $Z_{DD}=2(1-\gamma)$, and $\alpha^{2}=Z_{DD}^{-1}-\dfrac{1}{4}$.
Here the momentum $k_{*,1(2)}$ denotes the relative momentum of the corresponding pair in the rest frame.

The kernel of the scattering equation is denoted by $Z$ in three channels,
\begin{align}
Z=\, & \begin{pmatrix}Z_{1} & Z_{12}\\
Z_{21} & Z_2
\end{pmatrix}.
\end{align}
In principle, the potential $Z$ should include both particle exchange terms and contact terms. As we have discussed before, the proper contact term is introduced in the form of non-derivative coupling, i.e., $h_{0}$ in the $(DK) + D\to(DK)+ D$ channel. This is just one choice   to parameterize the 3-body force of the $DDK$ system. More explicitly, the 3-body force is reparameterized in the potential,
\begin{align}
\frac{H_{0}}{\Lambda^{2}}= & -\frac{h_{0}}{2\mu f_{0,DK}^{2}}.
\end{align} 
Additionally, the component $Z_1$ includes also the potential of $(DK)+D \to (DK)+D$ by exchanging the $K$ meson,  $Z_{12}$ includes the potential $(DK)+D \to (DD)+K$ by exchanging the $D$ meson and $Z_{21}$  inversely.  In a closed form, $Z_2$ is designed for the potential of $(DD)+K \to (DD)+K$.
However,  there is no one-particle exchange potential in this reaction and therefore, we are allowed to set $Z_2=0$ at the beginning.
Finally, we arrive at the elements of $Z$ which can be written as,
\begin{align}
Z_{1}(\mathbf{p},\mathbf{q};E)=\, & \frac{1}{p^{2}+q^{2}+2\gamma\mathbf{p}\mathbf{q}-2\mu E}+\frac{H_{0}}{\Lambda^{2}};\\
Z_{12}(\mathbf{p},\mathbf{q};E)=\, & \frac{1}{2(1-\gamma)p^{2}+q^{2}+2(1-\gamma)\mathbf{p}\mathbf{q}-2\mu E};\\
Z_{21}(\mathbf{p},\mathbf{q};E)=\, & Z_{12}(\mathbf{q},\mathbf{p};E).
\\
Z_2(\mathbf{p},\mathbf{q};E)=0
\end{align}

Corresponding to Eq.(\ref{eq:scattering-eq}), the finite volume amplitude $\mathcal{M}_{L}$ satisfies the following scattering equation,
\begin{align}
\mathcal{M}_{L}(\mathbf{p},\mathbf{q};E)= & Z(\mathbf{p},\mathbf{q};E)+\frac{4\pi}{L^{3}}\sum_{\mathbf{k}}^{\Lambda}Z(\mathbf{p},\mathbf{k};E)\tau_{L}(\mathbf{k};E)\mathcal{M}_{L}(\mathbf{k},\mathbf{q};E).\label{eq:finite-volume-equation}
\end{align}
$L$ is the spatial size of the cubic box, $E$ is thus the finite-volume
energy level. Here $\mathbf{p},\mathbf{q}$ and $\mathbf{k}$ are
the discretized relative dimer-spectator three-momenta, i.e., $\mathbf{p},\mathbf{q},\mathbf{k}\in\{2\pi\mathbf{n}/L|\mathbf{n}\in\mathbb{Z}^{3}\}$.
The 2-body finite-volume correction is encoded in $\tau_{L}$ as
\begin{align}
\tau_{L,1}(\mathbf{k};E)=\frac{1}{k_{*,1}\cot\delta_{DK}-S_{L}^{(DK)}(\mathbf{k};E)};\\
\tau_{L,2}(\mathbf{k};E)=\frac{2Z_{DD}}{k_{*,2}\cot\delta_{DD}-S_{L}^{(DD)}(\mathbf{k};E)},
\end{align}
where the finite volume corrections read,
\begin{align}
S_{L}^{(DK)}(\mathbf{k};E) & =\left(\frac{1}{L^{3}}\sum_{\mathbf{l}}-\text{PV}\int\frac{d^{3}l}{(2\pi)^{3}}\right)\frac{1}{(\mathbf{l}+\gamma\mathbf{k})^{2}-k_{1*}^{2}};\\
S_{L}^{(DD)}(\mathbf{k};E) & =\left(\frac{1}{L^{3}}\sum_{\mathbf{l}}-\text{PV}\int\frac{d^{3}l}{(2\pi)^{3}}\right)\frac{1}{(\mathbf{l}+\mathbf{k}/2)^{2}-k_{2*}^{2}}.
\end{align}
The 3-body force does not need to be modified in Eq.(\ref{eq:finite-volume-equation}) because it describes short-range interaction in the system which is not sensitive to the finite volume effect. 

\subsection{Quantization condition}

The energy levels on lattice are determined by the pole position of finite volume amplitude $\mathcal{M}_{L}$.
Since these energy levels are always discrete, the scattering equation can be reduced to a homogeneous equation.
Near the pole of $\mathcal{M}_{L}$ at $E_{*}$ , the amplitude can be written in the form of
\begin{align}
\mathcal{M}_{L}(\mathbf{p},\mathbf{q};E)= & \frac{\phi(\mathbf{p})\phi^{*}(\mathbf{q})}{E-E_{*}}+\text{reg. }.
\end{align}
So the homogeneous equation is
\begin{align}
\phi(\mathbf{p})= & \frac{4\pi}{L^{3}}\sum_{\mathbf{k}}^{\Lambda}Z(\mathbf{p},\mathbf{k})\tau_{L}(\mathbf{k})\phi(\mathbf{k}).
\end{align}
To be consistent with $Z$ and $\tau_{L}$, the factorized component $\phi$ also consists of  two channels, i.e., $\phi=(\phi_{1},\phi_{2})^{\text{T}}$.
Using the cubic symmetry projection of \cite{pang3}, we have the following equation in the $A_{1}^{+}$ irreducible representation($A_{1}^{+}$-irreps.),
\begin{align}
\phi_{r}=\, & \frac{4\pi}{L^{3}}\sum_{s}^{s_{\Lambda}}\vartheta_{s}\,Z_{rs}^{(A_{1}^{+})}\tau_{s}\phi_{s}.
\end{align}
The indices $r,s$ denote the  shells to which the discrete momenta belong. Correspondingly,  $s_{\Lambda}$ is the cutoff shell.
We call $\vartheta_{s}$ as the number of momenta in shell $s$.
The projected potential and dimer propagator are
\begin{align}
Z_{rs}^{(A_{1}^{+})}=\, & \frac{1}{48}\sum_{g\in G}Z(g\mathbf{p}_{0}^{(r)},\mathbf{k}_{0}^{(s)}),\quad\tau_{s}=\tau_{L}(\mathbf{k}_{0}^{(s)}),
\end{align}
where $\mathbf{p}_{0}^{(r)},\mathbf{k}_{0}^{(s)}$ are reference momenta of $r,s$ shells respectively. $g$ is octahedral transformation of the group $G$.
At last, we obtain the 3-body quantization condition of the $DDK$ system in the $A_{1}^{+}$-irreps.,
\begin{align}
\det\left[\delta_{rs}\begin{pmatrix}\tau_{L,1}^{-1}(s)\\
 & \tau_{L,1}^{-1}(s)
\end{pmatrix}-\frac{4\pi}{L^{3}}\vartheta_{s}\begin{pmatrix}Z_{1,rs}^{(A_{1}^{+})} &  & Z_{12,rs}^{(A_{1}^{+})}\\
\\
Z_{21,rs}^{(A_{1}^{+})} &  & 0
\end{pmatrix}\right] & =0.\label{eq:quantization-condition}
\end{align}

\section{\label{sec:Result}Results and Discussions}

\subsection{2-body part}

In order to resolve the 3-body dynamics, it is necessary to calculate the 2-body scattering amplitude firstly.
Therefore, we solve the Lippmann-Schwinger\corwu{}{(LS)} equation based on the $DK$ and $DD$ potential.
The equation is written down as,
\begin{align}
T_{\ell}(p,q)= & V_{\ell}(p,q)+\frac{1}{2\pi^{2}}\int_{0}^{\infty}k^{2}dkV_{\ell}(p,k)G(k)T_{\ell}(k,q)
\end{align}
Here $T_{\ell}$ is the partial-wave scattering amplitude.
The corresponding potential is projected as
\begin{align}
V_{\ell}(p,k)= & \frac{1}{2}\int_{-1}^{1}d\cos\theta\,P_{\ell}(\cos\theta)\,V(|\mathbf{p}-\mathbf{k}|),\quad\text{with }\mathbf{p}\cdot\mathbf{k}=pk\cos\theta.
\end{align}
We can calculate the potential in momentum space $V(|\mathbf{p}-\mathbf{k}|)$ by Fourier transformation in the form of,
\begin{align}
V(|\mathbf{p}-\mathbf{k}|)= & \int d^{3}x\,\tilde{V}(\mathbf{x})e^{-i(\mathbf{p}-\mathbf{k})\mathbf{x}}\,.
\end{align}
To complete the LS equation, the 2-body propagator takes the form of \footnote{For the $DD$ system, we also need to consider the additional symmetry factor,
$1/2$.} $G(k)=1/\left(E-\dfrac{k^{2}}{2\mu}+i\epsilon\right) $.

The 2-body interaction inside the $DDK$ system is described in terms of  effective potentials in \cite{Wu:2019vsy}.
We introduce the $DK$ interacting subsystem by considering a contact-range effective field theory where
the leading order contribution is
\begin{align}
\tilde{V}_{DK}(r;R_{c})= & C_{L}^{'}e^{-(r/R_{c})^{2}}.
\end{align}
Here the interaction coupling constant takes the value of $C_{L}^{'}=-320.1\,\text{MeV}$ and the  typical interaction length is $R_{c}=1\,\text{fm}$.
At the same time, the $DD$ interaction is described  by the one boson exchange (OBE) potential involving $\sigma$, $\rho$ and $\omega$ mesons \cite{Liu:2019stu}:
\begin{align}
V_{DD}(\mathbf{q})= & \sum_{V=\sigma,\rho,\omega}C_{\text{iso.}}(V)\frac{g_{V}^{2}}{\mathbf{q}^{2}+m_{V}^{2}}\left(\frac{\tilde{\Lambda}^{2}-m_{V}^{2}}{\mathbf{q}^{2}+\tilde{\Lambda}^{2}-q_{0}^{2}}\right)^{2}.\label{eq:VDD}
\end{align}
Here the isospin factors are $C_{\text{iso.}}(\sigma)=-1$ , $C_{\text{iso.}}(\rho)=1$,  and $C_{\text{iso.}}(\omega)=+1$ .
$m_{V}$ is the mass of the vector meson.
The couplings are chosen as $g_{\sigma}=3.4$, $g_{\rho}=g_{\omega}=2.6$ with the corresponding cutoff parameter $\tilde{\Lambda}=1\,\text{GeV}$.

\begin{figure}
\begin{centering}
\includegraphics[scale=0.2]{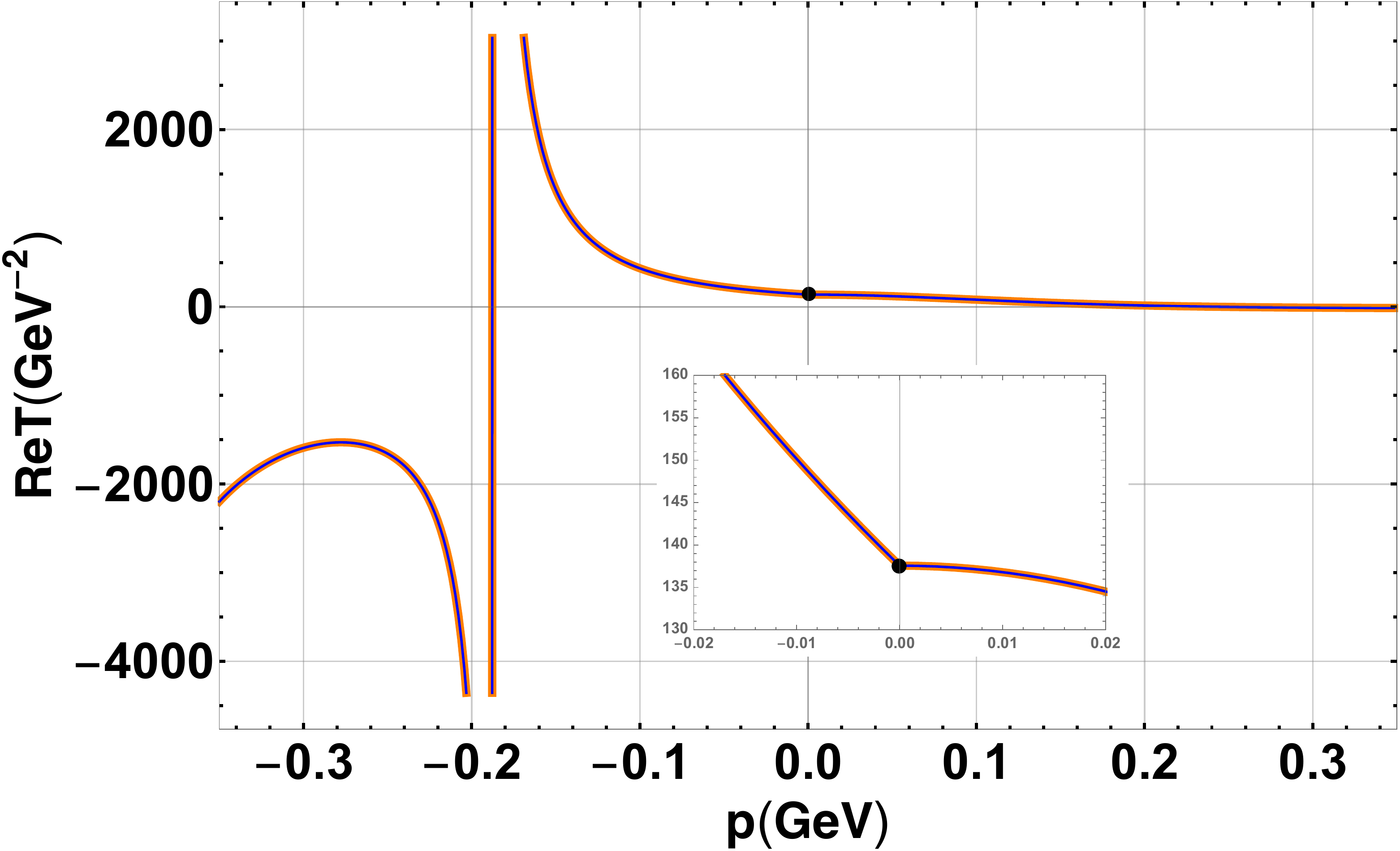}$\qquad$$\qquad$$\qquad$\includegraphics[scale=0.2]{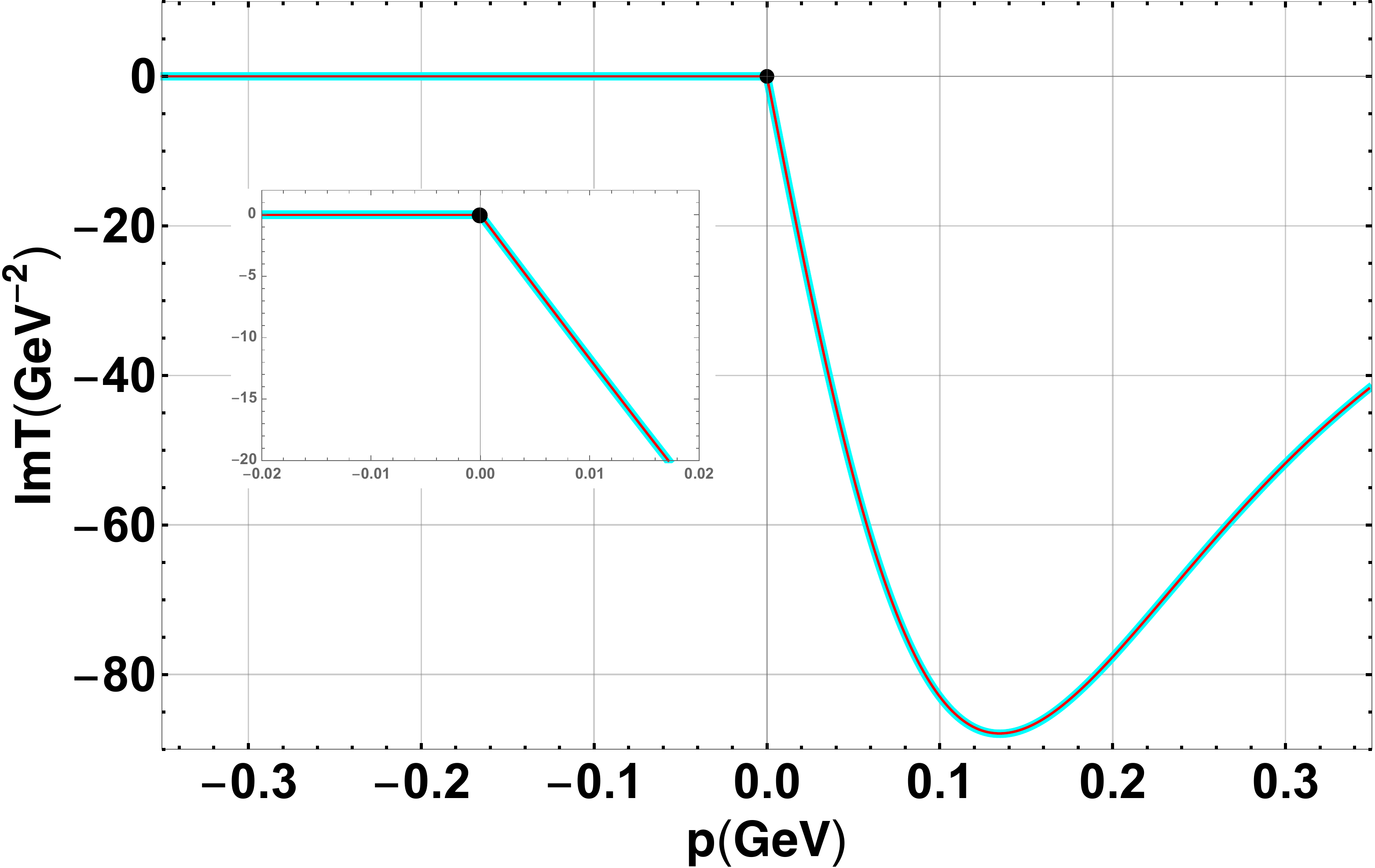}
\par\end{centering}
\caption{\label{fig:-scattering-amplitudes-DK}$DK$ scattering amplitudes.}
\end{figure}
\begin{figure}
\begin{centering}
\includegraphics[scale=0.2]{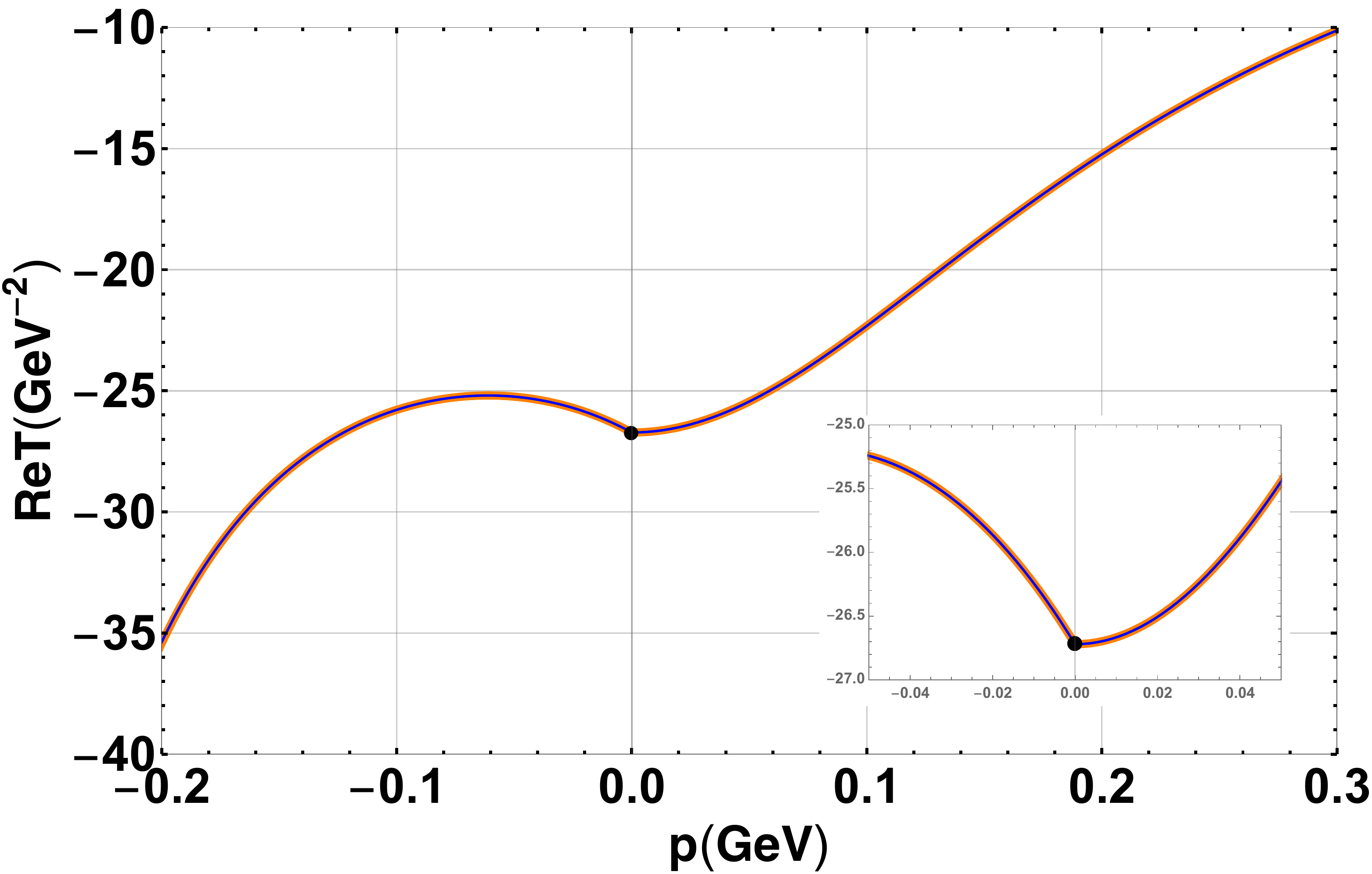}$\qquad$$\qquad$$\qquad$\includegraphics[scale=0.2]{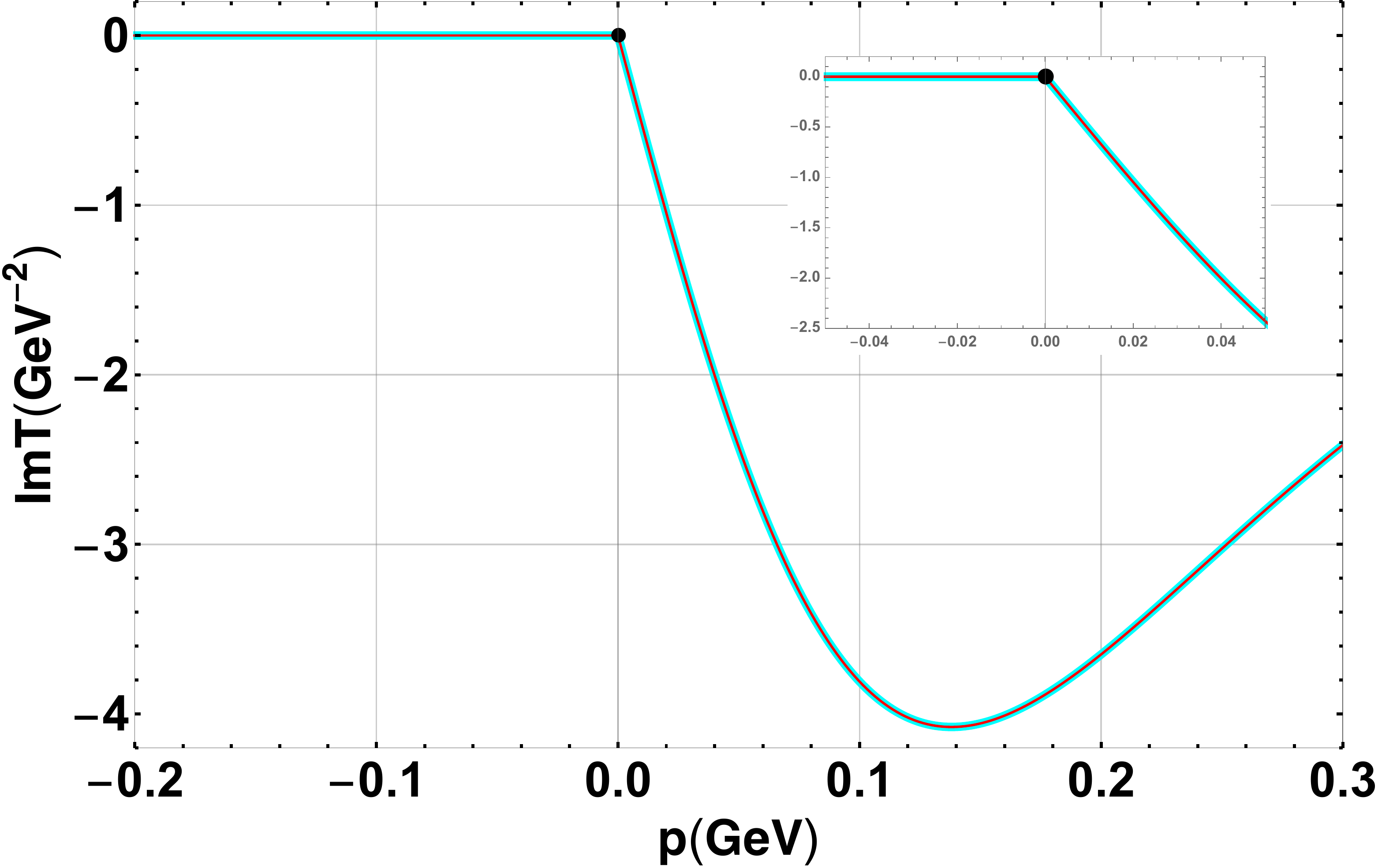}
\par\end{centering}
\caption{\label{fig:-scattering-amplitudes-DD}$DD$ scattering amplitudes.}
\end{figure}

By solving the $S$-wave LS equation with the $DK$ and $DD$ interaction potential, we can obtain the 2-body scattering amplitudes (see Figs. \ref{fig:-scattering-amplitudes-DK} and \ref{fig:-scattering-amplitudes-DD}).
Here, the amplitude below the threshold is the analytic continuation since it is required in the 3-body calculation.
We need to reexpress the amplitude in temrs of momentum expansion as follows,
\begin{align}
T_{DK}(E)= & -\frac{2\pi}{\mu}\begin{cases}
\dfrac{1}{-a_{DK}^{-1}+\dfrac{1}{2}r_{DK}p^{2}-ip}, & p=\sqrt{+2\mu E},\quad E>0;\\
\\
\dfrac{C_{DK}}{\kappa-\kappa_{DK}}+R_{DK}+R_{DK}^{(1)}\kappa+R_{DK}^{(2)}\kappa^2, & \kappa=\sqrt{-2\mu E},\quad E<0,
\end{cases}
\end{align}
and
\begin{align}
T_{DD}(E)= & -\frac{8\pi}{m_{D}}\begin{cases}
\dfrac{1}{-a_{DD}^{-1}+\dfrac{1}{2}r_{DD}p^{2}-ip}, & p=\sqrt{+m_{D}E},\quad E>0;\\
\\
\dfrac{C_{DD}}{\kappa-\kappa_{DD}}+R_{DD}+R_{DD}^{(1)}\kappa+R_{DD}^{(2)}\kappa^{2}, & \kappa=\sqrt{-m_{D}E},\quad E<0.
\end{cases}
\end{align}
\corwu{We are using the reduced mass of $DK$, $\mu=m_{D}m_{k}/(m_{D}+m_{K})$.}{In the above definitions, $\mu$ is the reduced mass of $DK$.}
The corresponding expansion parameters are listed in Table \ref{tab:T-DK-para-fm},
which are all fitted from the amplitude obtained from the LS equation.
One should note that above the threshold, we use the normal effective range expansion, while the expression below the threshold is different because the quantity $p\cot\delta-i p$  induces additional spurious poles far below threshold, where $p$ is purely imaginary. These spurious poles emerge at the hard scale of the theory, i.e., outside the range, where the effective field theory is applicable. In order to protect the unitarity in our formalism, we give an alternative expression and
we can explicitly see that in the proper expression below threshold, there is only one physical pole for the $DK$ system (no pole for the $DD$ system),
\begin{align}
m_{D}+m_{K}-\frac{\kappa_{DK}^{2}}{2\mu}\simeq & \,2318.
\end{align}
This pole is interpreted as the $DK$ bound state, $D_{s0}^{*}(2317)$ which has been supported by both lattice QCD \cite{MartinezTorres:2011pr,Guo:2018kno,Liu:2012zya,Guo:2018ocg,Guo:2018tjx,Mohler:2013rwa,Lang:2014yfa,Torres:2014vna,Bali:2017pdv}
and phenomenological studies \cite{Kolomeitsev:2003ac,Hofmann:2003je,Guo:2006fu,Gamermann:2006nm,Guo:2008gp,Guo:2009ct,Cleven:2010aw,Yao:2015qia,Guo:2015dha,Albaladejo:2016lbb,Du:2017ttu,Altenbuchinger:2013vwa,Altenbuchinger:2013gaa,Albaladejo:2018mhb,Geng:2010vw,Wang:2012bu,Liu:2009uz}.
Based on the $DD$ interaction potential, Eq.(\ref{eq:VDD}) referring to \cite{Wu:2019vsy,Liu:2019stu}, we also find a virtual state in the $DD$ system, $\kappa_{DD}\simeq-195\text{MeV}$.

\begin{table}
\begin{centering}
\begin{tabular}{c|c|c|c|c|c|c|c}
\hline\hline
 & \multirow{3}{2cm}{$a$ (fm)} & \multirow{3}{1.5cm}{$r$ (fm)} & \multirow{3}{1.7cm}{$\kappa$ (MeV)} & \multirow{3}{1.3cm}{$C$ } & \multirow{3}{2cm}{$R$ (MeV$^{-1}$)} & \multirow{3}{2cm}{$R^{(1)}$ (MeV$^{-1}$)} & \multirow{3}{2cm}{$R^{(2)}$ (MeV$^{-1}$)}\tabularnewline
 &  &  &  &  &  &  & \tabularnewline
 &  &  &  &  &  &  & \tabularnewline
\hline
\multirow{2}{1.2cm}{$DK$} & \multirow{2}{2cm}{$1.683$ $\big(1.58_{-0.17}^{+0.22}\big)$ } & \multirow{2}{1.5cm}{$0.791$} & \multirow{2}{1.7cm}{$187.795$} & \multirow{2}{1.3cm}{$3.881$} & \multirow{2}{2cm}{$0.0121$} & \multirow{2}{2cm}{$3.73\times10^{-5}$} & \multirow{2}{2cm}{$1.11\times10^{-7}$}\tabularnewline
 &  &  &  &  &  &  & \tabularnewline
\hline
\multirow{2}{1.2cm}{$DD$} & \multirow{2}{2cm}{$-0.392$ $\big(-0.4_{-0.2}^{+0.1}\big)$} & \multirow{2}{1.5cm}{$3.236$} & \multirow{2}{1.7cm}{$-195.166$} & \multirow{2}{1.3cm}{$0.243$} & \multirow{2}{2cm}{$7.43\times10^{-4}$} & \multirow{2}{2cm}{$2.43\times10^{-6}$} & \multirow{2}{2cm}{$7.52\times10^{-9}$}\tabularnewline
 &  &  &  &  &  &  & \tabularnewline
\hline\hline
\end{tabular}
\par\end{centering}
\caption{\label{tab:T-DK-para-fm}Parameters of the scattering amplitude up to NLO.
The scattering length of $DK$ is consistent with the result of \cite{Guo:2018tjx}
result in bracket and also other lattice simulation results in \cite{Yao:2015qia,Liu:2012zya,Mohler:2013rwa,Lang:2014yfa,Torres:2014vna,Bali:2017pdv}.
Note that the sign convension of the scattering length in lattice simulations
is opposite to that of our phenomenological studies. On the other hand, the scattering length of
$DD$ is consistent the result of \cite{Liu:2019stu} in bracket as
well. }
\end{table}
\subsection{3-body part}

\subsubsection{3-body system in infinite volume}

\begin{figure}
\begin{centering}
\includegraphics[scale=0.25]{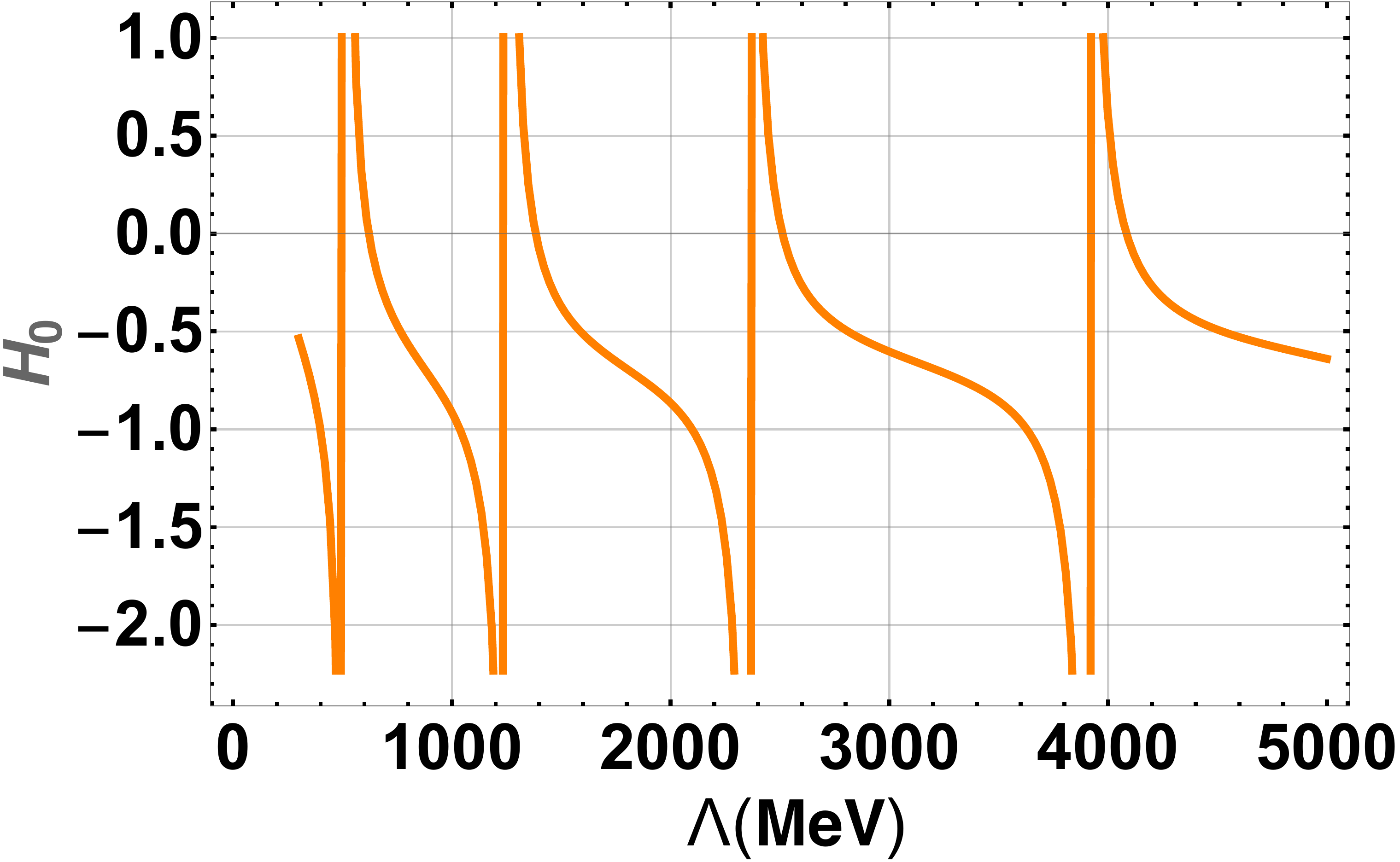}$\qquad\quad$\includegraphics[scale=0.25]{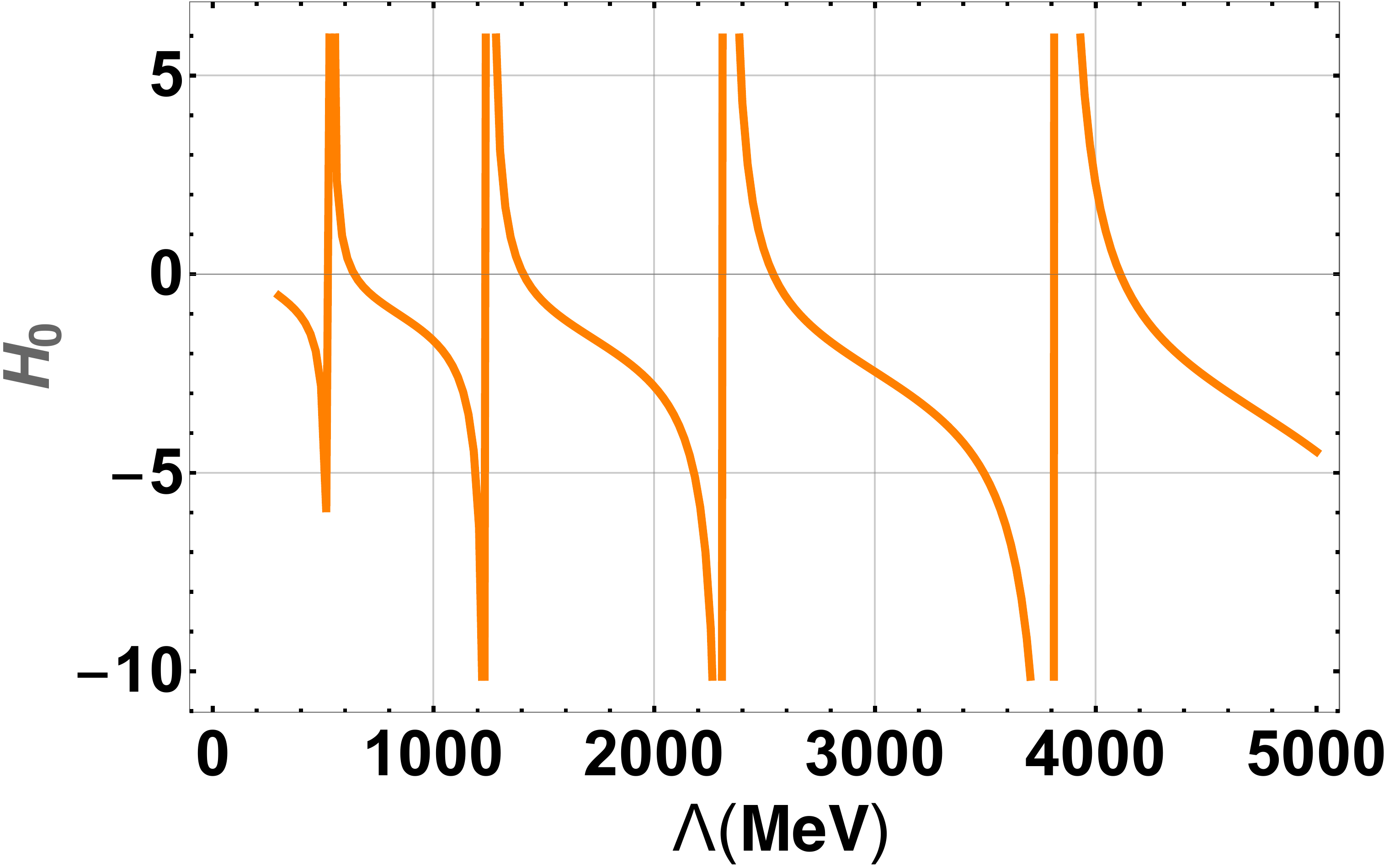}
\par\end{centering}
\caption{\label{fig:Three-body-force}$DDK$ three-body force as a function of the cutoff. Left:
without the $DD$ interaction;  Right: with the meson exchange interaction between
the $D$ mesons. }
\end{figure}

We introduced a 3-body force to describe the short range interaction in the $DDK$ system.
It turns out that the 3-body force $H_{0}$ should be running periodically as shown in Fig. \ref{fig:Three-body-force}.
Consequently, the solution of the particle-dimer scattering equation (\ref{eq:scattering-eq}) shows that there is a 3-body bound state pole in the  $DDK$ system which is consistent with the prediction of \cite{Wu:2019vsy} (see Table
\ref{tab:DDK-body-bound-state}).
Actually, the behavior of the 3-body force as shown in Fig. \ref{fig:Three-body-force} is determined from the binding energy of the 3-body bound state which can be obtained in either the phenomenological study - that we refer to - or an experiment measurement.  This relationship between $H_0$ and $\Lambda$ can keep our predictions totally cutoff independent.
\footnote{Sometimes,  one can choose a proper value for the cutoff, $\Lambda$, so that $H_0$ is typically small, then this 3-body force will be effectively suppressed  in the corresponding model.}

\begin{table}
\begin{centering}
\begin{tabular}{c|c|c|c|c}
\hline\hline
\multirow{2}{2cm}{$C_{S}^{'}$ (MeV)} & \multirow{2}{2cm}{$C_{L}^{'}$ (MeV)} & \multirow{2}{2cm}{$E_{2}$ (MeV)} & \multirow{2}{2.2cm}{$E_{3}$ (only $DK$)} & \multirow{2}{3.8cm}{$E_{3}$ (both $DK$ and $DD$)}\tabularnewline
 &  &  &  & \tabularnewline
\hline
\multicolumn{5}{c}{\multirow{2}{12cm}{$R_{L}=1\text{fm},R_{S}=0.5\text{fm}$}}\tabularnewline
\multicolumn{5}{c}{}\tabularnewline
\hline
\multirow{2}{2cm}{$0$} & \multirow{2}{2cm}{$-320.1$} & \multirow{2}{2cm}{$-45.0$} & \multirow{2}{2.2cm}{$-65.8$} & \multirow{2}{3.8cm}{$-71.2$}\tabularnewline
 &  &  &  & \tabularnewline
\hline\hline
\end{tabular}
\par\end{centering}
\caption{\label{tab:DDK-body-bound-state}The 3-body $DDK$ bound state 
predicted by \cite{Wu:2019vsy}. }

\end{table}

\subsubsection{3-body system in  finite volume}

\begin{figure}
\begin{centering}
\includegraphics[scale=0.25]{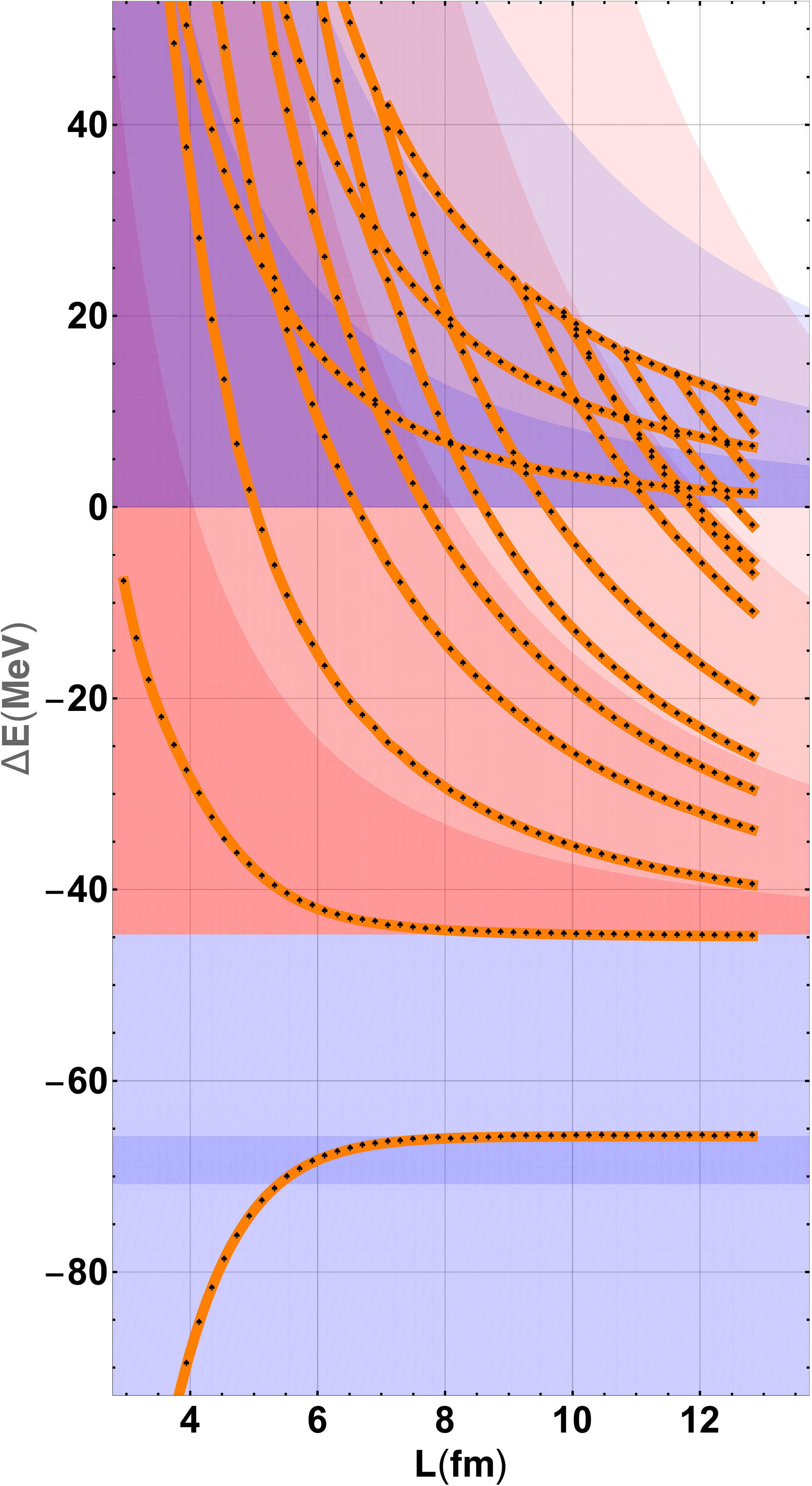}$\qquad\quad\qquad$\includegraphics[scale=0.25]{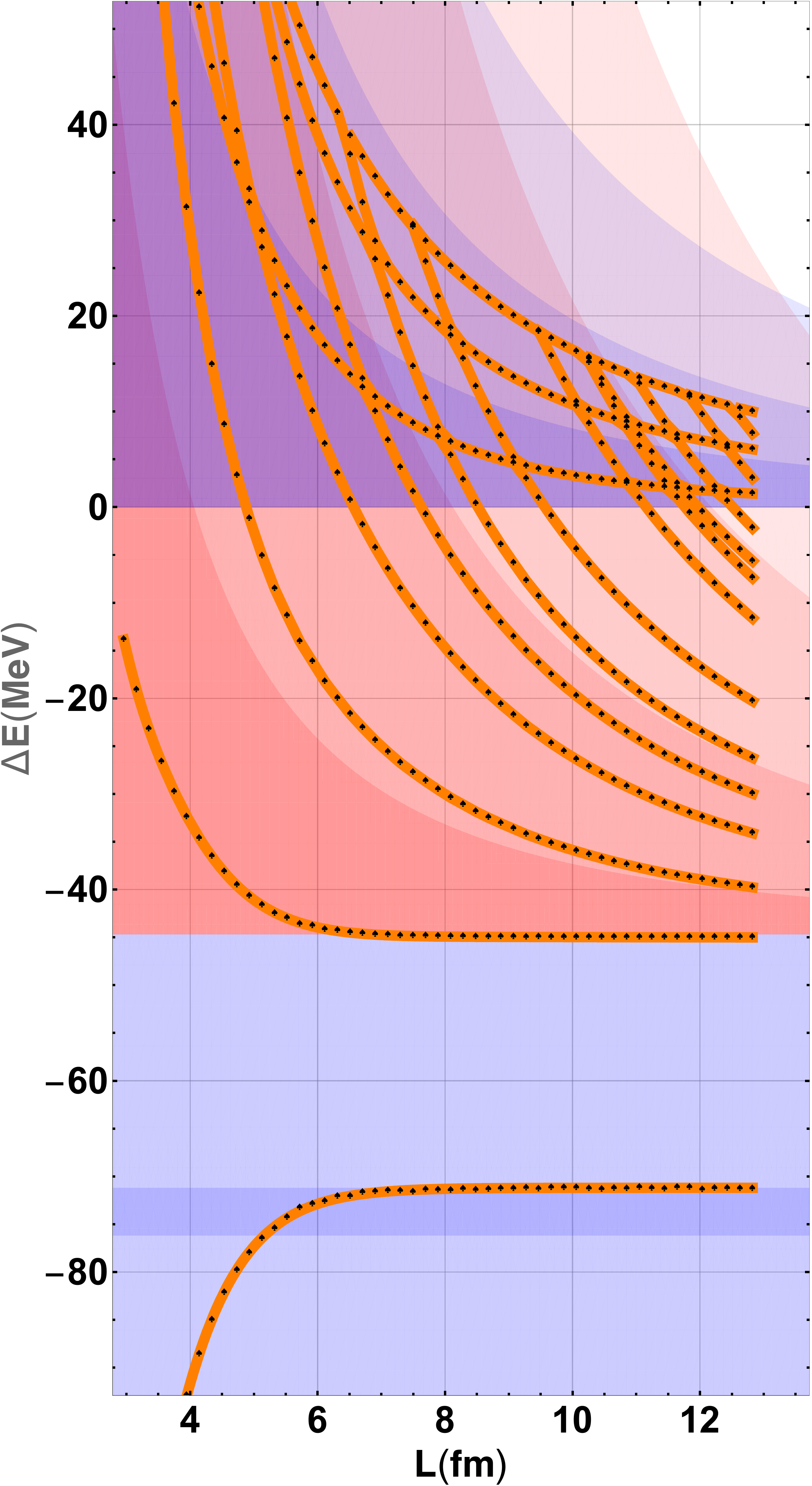}
\par\end{centering}
\caption{\label{fig:spectrum-121-122}$DDK$ states in finite volume. Left:
only the $DK$ interaction is considered. Right: both $DK$ and $DD$ interactions are taken into account. The upper blue regions indicate
the  case of 3  free particles in finite volume. The red regions indicate
the  case of free $D_{s0}^{*}(2317)$ and $D$. The lower blue regions
indicate the $DDK$ bound state below the $DD_{s0}^{*}(2317)$ threshold. }
\end{figure}

\begin{figure}
\begin{centering}
\includegraphics[scale=0.3]{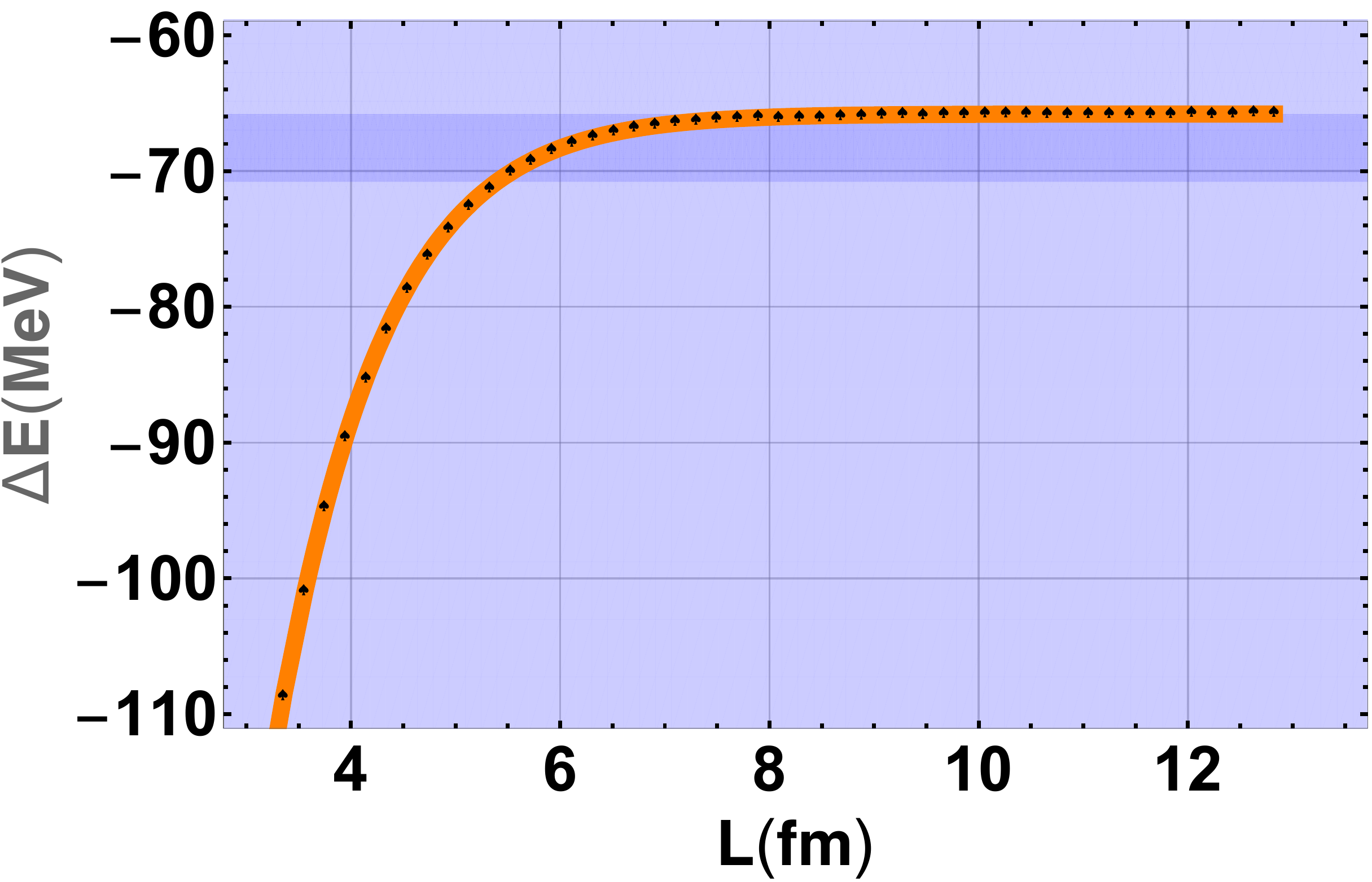}$\qquad\quad\qquad$\includegraphics[scale=0.3]{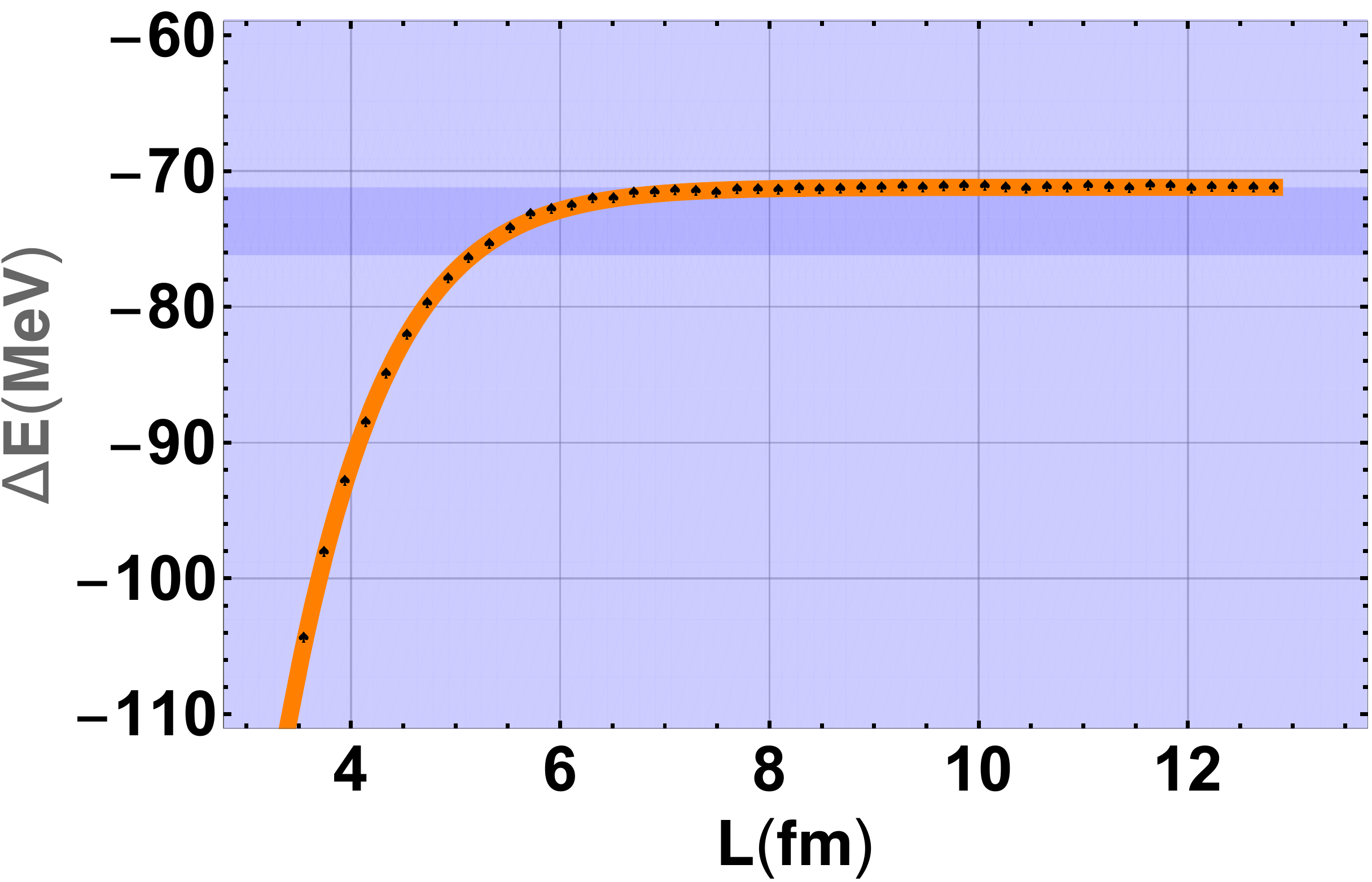}
\par\end{centering}
\caption{\label{fig:bound-state-121-122}$DDK$ bound states in finite volume. Left:
only the $DK$ interaction is considered. Right: both $DK$ and $DD$ interactions are taken into account. The deeper blue regions
indicate the $DDK$ bound state below the $DD_{s0}^{*}(2317)$ threshold. }
\end{figure}

Inputting the 3-body force $H_{0}(\Lambda)$ into the quantization condition (\ref{eq:quantization-condition}), we can obtain the lattice spectrum of the $DDK$ 3-body system, (see Fig. \ref{fig:spectrum-121-122}).
Two thresholds are presented.
The threshold at $\Delta E=0$ denotes the 3-body threshold which in fact means the energy at $m_{K}+2m_{D}$.
There is an additional 2-body threshold at $\Delta E=-45\,\text{MeV}$ which denotes the $DD_{s0}^{*}(2317)$ threshold.
This is the consequence of involving the $DK$ 2-body bound state in our formalism.
The existence of two thresholds is consistent with our previous toy model \cite{pang1,pang2,pang3,pang4}.
Therefore, we also find the avoided level crossing between spectra tending to the 3-body threshold and the 2-body threshold. The exclusion of the additional threshold implies that the $DD$ system is unbound.
Otherwise, there should be a third threshold which is related to the $DD$ bound state and the spectating $K$ meson.
In  summary, the 3-body scattering states of $DDK$ live above the 3-body threshold while the 2-body scattering states of $DD_{s0}^{*}(2317)$ live above the 2-body threshold.
Finally, below the 2-body $DD_{s0}^{*}(2317)$ threshold, the 3-body bound state in finite volume is found as well by checking the lowest energy level of Fig. \ref{fig:spectrum-121-122}.
A closer look at Fig. \ref{fig:bound-state-121-122}) reveals that the finite volume energy level exhibits an exponential behavior tending to the bound state ($-65.8\,\text{MeV}$ without $DD$ interaction and $-71.2\,\text{MeV}$ with both $DK$ and $DD$ interaction) in  infinite volume as predicted by \cite{Wu:2019vsy}.

%

\subsection{Comparing the $O(p^{2})$ results with the $O(p^{0})$ and $O(p^{4})$ results}

\begin{figure}
\begin{centering}
\includegraphics[scale=0.4]{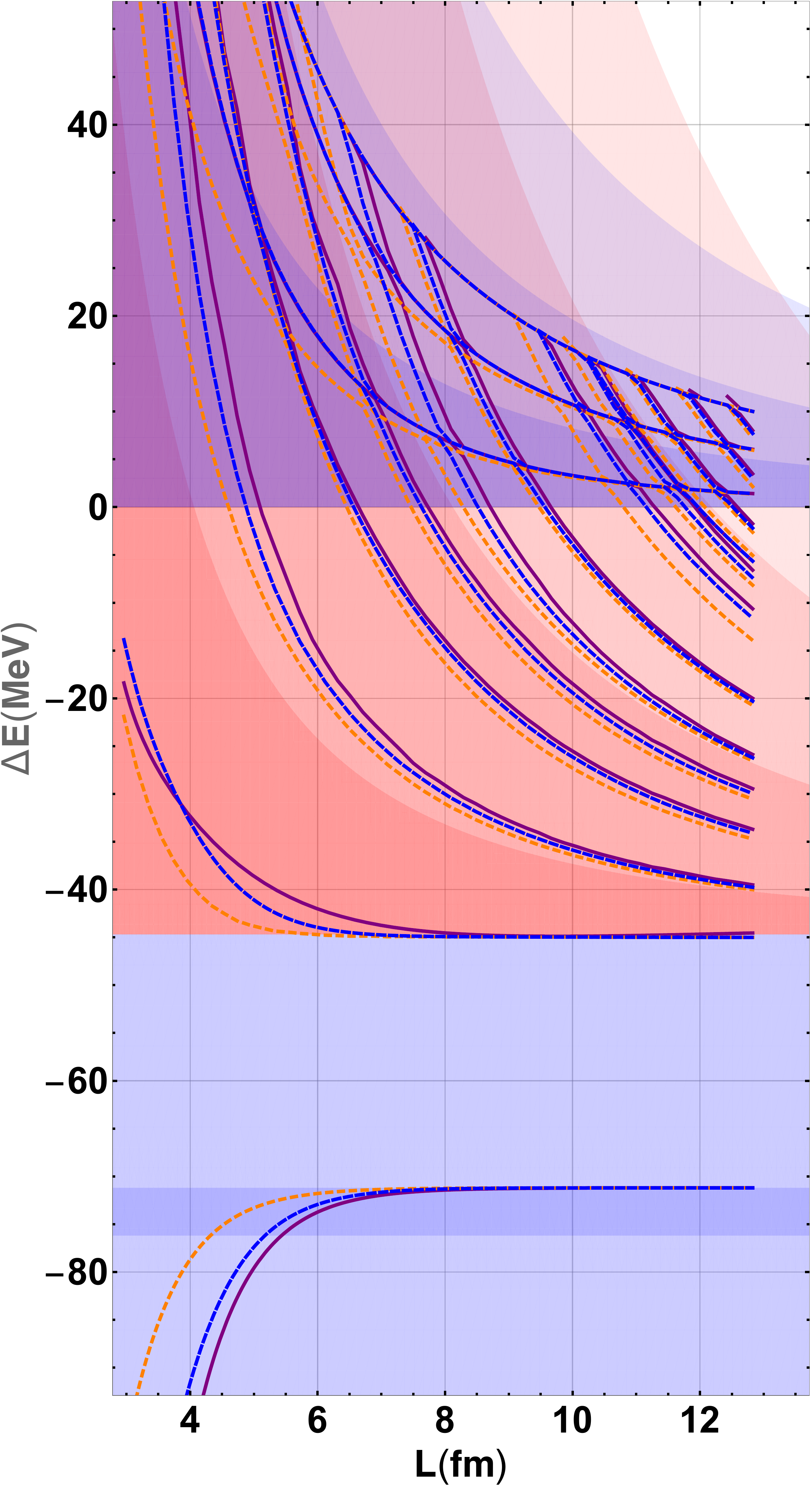}
\par\end{centering}
\caption{\label{fig:DDK-order-compare}$DDK$ states in  finite volume. The
orange curves are calculated at $O(p^{0})$, the blue curves at $O(p^{2})$,
and the purple curves at $O(p^{4})$. }
\end{figure}
The 2-body interaction involving $DK$ and $DD$ are resolved at $O(p^{2})$.
We also perform the calculation by truncating the 2-body part at $O(p^{0})$ and expanding it up to $O(p^{4})$.
At $O(p^{4})$, the parameters are listed in Table \ref{tab:T-DK-para-1}.
We find that the lattice spectrum of the $DDK$ 3-body system turns to be stable in a general profile (see Fig. \ref{fig:DDK-order-compare}).

\begin{figure}
\begin{centering}
\includegraphics[scale=0.4]{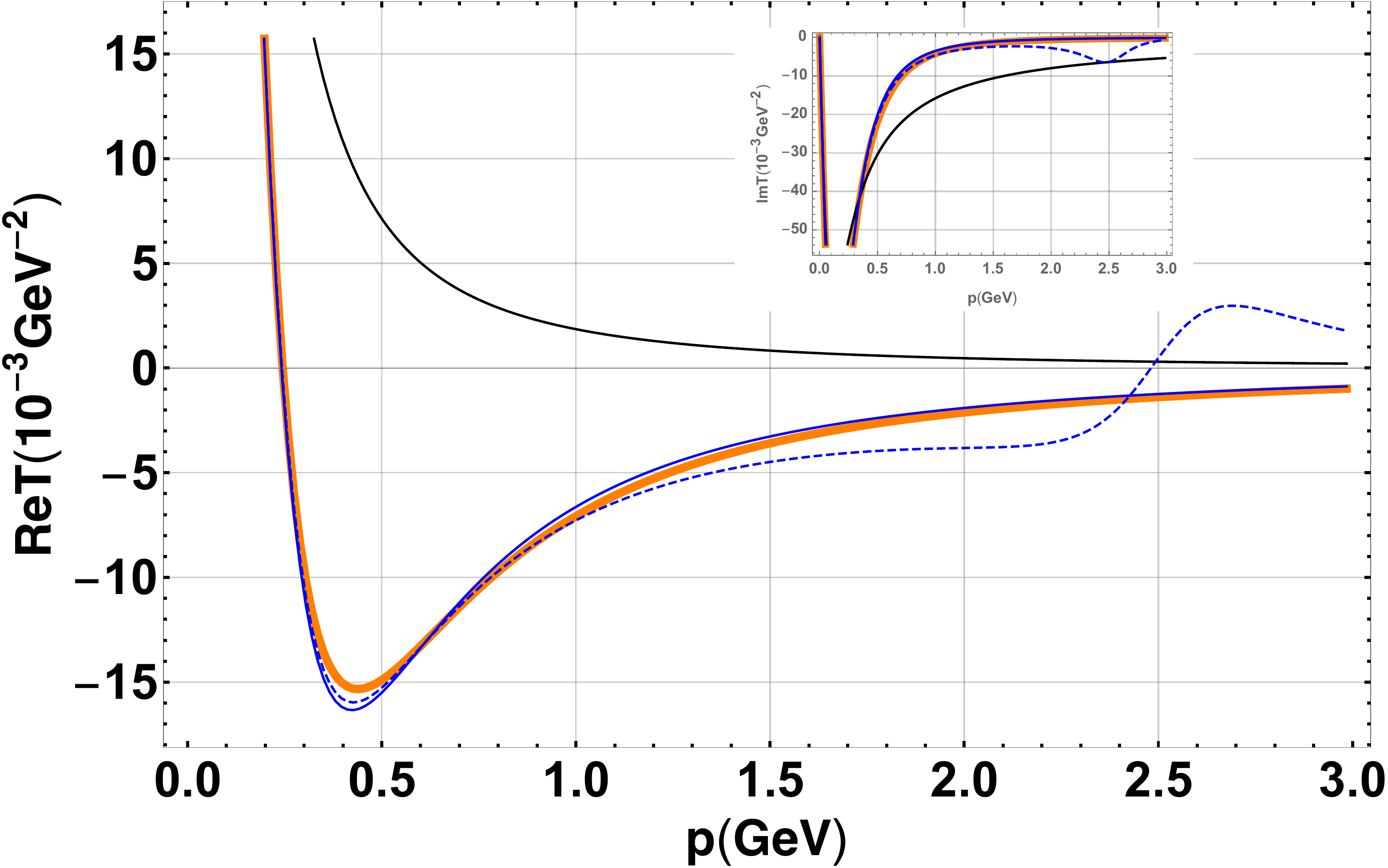}
\par\end{centering}
\caption{\label{fig:2-body-phase-shift}Two-body amplitude and the fitting up
to $O(p^{0})$, $O(p^{2})$ and $O(p^{4})$. The orange line denotes
the numerical solution of the LS equation. The black line is at $O(p^{0})$,
the blue line is at $O(p^{2})$, and the blue dashed line is at $O(p^{4})$. }
\end{figure}

\begin{table}
\begin{centering}
\begin{tabular}{c|c|c|c}
\hline\hline
 & \multirow{2}{3cm}{$P_{4}$ (MeV$^{-3}$)} & \multirow{2}{3cm}{$R^{(3)}$ (MeV$^{-4}$)} & \multirow{2}{3cm}{$R^{(4)}$ (MeV$^{-5}$)}\tabularnewline
 &  &  & \tabularnewline
\hline
\multirow{2}{1cm}{$DK$} & \multirow{2}{3cm}{$-3.228\times10^{-10}$} & \multirow{2}{3cm}{$3.17\times10^{-10}$} & \multirow{2}{3cm}{$8.49\times10^{-13}$}\tabularnewline
 &  &  & \tabularnewline
\hline
\multirow{2}{1cm}{$DD$} & \multirow{2}{3cm}{$2.49\times10^{-9}$} & \multirow{2}{3cm}{$2.33\times10^{-11}$} & \multirow{2}{3cm}{$7.27\times10^{-14}$}\tabularnewline
 &  &  & \tabularnewline
\hline\hline
\end{tabular}
\par\end{centering}
\caption{\label{tab:T-DK-para-1}Parametes of the scattering amplitude at N$^{2}$LO.
The effective range expansion at N$^{2}$LO reads $p\cot\delta=-a^{-1}+\dfrac{1}{2}rp^{2}+P_{4}p^{4}$. }
\end{table}
The $DK$ 2-body amplitude fitted up to different orders are shown in Fig. \ref{fig:2-body-phase-shift}.
The higher order fitting, i.e., $O(p^{4})$ expansion, gives better precision near the threshold, however, the
part far away presents a large discrepancy.
The 3-body spectrum turns out to depend on the 2-body phase shifts both near and far away from threshold.
We do not mean that the final spectrum is strongly related to the high-momentum dynamics, but it is fair to argue that up to very high order, the large discrepancy far away from threshold is able to disturb the result competing with the convergence near threshold. 
Since up to $O(p^{2})$, the result has stabilized, it is reasonable to stick to
the result at $O(p^{2})$.

For the different expansion orders of the 2-body interaction, the running of the 3-body force also shows different behavior.
Since here the 3-body force is in fact a counter-term,  the strength of this term will be sensitive to the precision of the 2-body interaction.

\section{\label{sec:Discussion}Summary}

We have successfully produced the 3-body bound state of $DDK$ on lattice (see Fig. \ref{fig:spectrum-121-122}). Our calculation is based on effective field theory referring to the phenomenological setup in \cite{Wu:2019vsy,Liu:2019stu}.
In the infinite volume limit, it tends to the realistic energy predicted by \cite{Wu:2019vsy}.
The energy level of spectrum for the bound state of three bodies shows the exponential behavior which is consistent with many previous calculations of Efimov states on lattice \cite{Meissner:2014dea,pang1,pang3}.
The prediction for the lattice spectrum of bound state can be tested by future lattice simulations. We can find whether lattice QCD will generate this stable 3-body bound state and further more, we are allowed to use lattice spectrum to constrain 3-body force $H_0$ which is a very important LEC in the effective field theory.  This cross-check presents a connection between phenomenological methods and lattice calculations on the 3-body
dynamics, thus supplies an ab initio way to study the $DDK$ system. 

In the lattice spectrum, there exist energy levels which tend to the $DD_{s0}^{*}$ threshold in the infinite volume limit. This is the consequence of the 2-body bound state $D_{s0}^{*}(2317)$.
In our formalism, we assume that the $D_{s0}^{*}(2317)$ is a $D-K$ molecule.
This will influence deeply the 3-body dynamics. That is to say,  we can find the avoided level crossing in the spectrum. In other words, 3-body lattice simulations are able to 
confirm the existences of $D_{s0}^{*}(2317)$ and its behavior in the finite volume of 3-body system.
%
Since the $DD$ system is unbound, we do not see any energy level which belongs to the pattern of $(DD)-K$. We believe that future lattice simulations of $DDK$ will open a new window to reveal these dynamics in the charm
sector.

Our prediction for the $DDK$ 3-body lattice spectrum is derived from the  non-relativistic effective field theory.
It is reasonable because the calculation is carried out near the threshold.
In order to justify this, we have checked the stability of the spectrum by including the 2-body dynamics at $O(p^{0})$, $O(p^{2})$ and $O(p^{4})$.
It shows that the 3-body spectrum is not sensitive to the physics in the high momentum region
of the relevant particles.
Higher order calculations in principle can be done, but are  unnecessary because when the order is
taken to too high, the effective range expansion which we have used in the 2-body matching will break the precision of the 3-body result.
So the result up to $O(p^{2})$ are already good enough.

The lattice spectrum below the $DD_{s0}^{*}$ threshold can give more detail to study the physical picture of the  $DDK$ bound state as we have done in the toy model \cite{pang3}.
The analytic expression for the 3-body ground state in \cite{pang4} can also be derived for the $DDK$ system and compared with the corresponding numerical result.
Additionally, the work can be improved by introducing a 3-body force beyond non-derivative coupling.
This is also a relevant topic for the cross-check with future lattice simulations.


\begin{acknowledgments}
The authors thank  Meng-Lin Du, Li Ma and Akaki Rusetsky for useful discussions. Jia-Jun Wu acknowledges the support from the Fundamental Research Funds for the Central Universities. Li-Sheng Geng is partly supported by the National Natural Science Foundation of China under Grants Nos.11735003, 11975041, and 11961141004.

\end{acknowledgments}

\bibliographystyle{unsrt}

\end{document}